\documentclass[11pt,a4paper]{article}
\usepackage{jcappub}
\usepackage{amsmath}
\usepackage{multirow}
\usepackage{graphicx}
\usepackage{bbm}
\usepackage{amssymb}
\usepackage{booktabs}

\title{Constraints on $\Lambda(t)$-cosmology with power law interacting dark sectors}
\author{Vincent Poitras}
\affiliation{McGill University, Department of Physics,\\
3600 University Street, Montreal, Canada, H3A 2T8.}
\emailAdd{poitrasv@physics.mcgill.ca}
\abstract{
Motivated by the cosmological constant and the coincidence problems, we consider a cosmological model where the cosmological constant $\Lambda_0$ is replaced by a cosmological term $\Lambda(t)$ which is allowed to vary in time. More specifically, we are considering that this dark energy term interacts with dark matter through the phenomenological decay law $\dot{\rho}_{\Lambda}=-Q\rho_{\Lambda}^{n}$.  We have constrained the model for the range $n\in[0,10]$ using various observational data (SNeIa, GRB, CMB, BAO, OHD), emphasizing the case where $n=3/2$. This case is the only one where the late-time value for the ratio of dark energy density and matter energy density $\rho_{\Lambda}/\rho_{m}$ is constant, which could provide an interesting explanation to the coincidence problem. We obtain strong limits on the model parameters which however exclude the region where the coincidence or the cosmological constant problems are significantly ameliorated.
}
\arxivnumber{1205.6766}
\begin{document}
  \maketitle 

  \section{Introduction}
It is now widely accepted by the cosmological community that our Universe is currently experiencing a phase of accelerated expansion.
The first significant evidences came from the distance measurements of type Ia supernovae (SNeIa) \cite{SNE1,SNE2,SNE3} (see \cite{SNE4,SNE5} for an update) which has been later confirmed, among others, by the measurement of the anisotropies of the cosmic microwave background (CMB) spectrum by the Wilkinson Microwave Anisotropy Probe (WMAP) \cite{CMB1,CMB2,CMB3,CMB4} and by the measurement of the baryon acoustic oscillations (BAO) in the Sloan Sky Digital Survey (SDSS) luminous galaxy sample \cite{BAO1,BAO2}.

Normal matter satisfies the strong energy condition ($w=p/\rho\geq0$) and cannot drive the accelerated expansion of the Universe. Hence, the recourse to a cosmic fluid, known as \emph{dark energy}, characterized by a negative pressure  ($w_x<0$), is usually made to explain the current accelerating expansion of the Universe. Alternatively to dark energy, other approaches such as modified gravity \cite{ModG1,ModG2,ModG3} and inhomogeneous cosmology \cite{Inhomo1,Inhomo2}, have been proposed to explain the (apparent) cosmic acceleration. 

The  simplest form  of dark energy is provided by a cosmological constant, which has led to the development of the $\Lambda$CDM model. In this model, the Universe is composed of, in addition to ordinary matter (radiation, $w_r=1/3$, baryon, $w_b=0$), a pressureless cold dark matter fluid ($w_c=0$) and a cosmological constant $\Lambda_0$ ($w_x=w_{\Lambda}=-1$).  The model provides a reasonably good fit to the current cosmological data but is however is plagued by two serious theoretical difficulties. One of them is the  \emph{cosmological constant problem}; the observed value of the cosmological constant is $\sim$120 orders of magnitude smaller than what is expected from theoric computations \cite{Weinberg}. The other one  relies on the observation that the values of the matter energy density ($\rho_m=\rho_b+\rho_c$) and of the dark energy density are currently of the same order of magnitude. However, since the former is diluted proportionally to the volume  of the Universe as it is expanding ($\rho_m\propto a^{-3}$) while the later remains constant ($\rho_{\Lambda}=\rho_{\Lambda_0}$), the period of time during which  $\rho_m/\rho_{\Lambda}\sim\mathcal{O}(1)$  is predicted to be very narrow in comparison to the whole Universe history. To currently lie in this narrow period, a fine tuning of the initial conditions of the model is needed. This is the essence of the \emph{coincidence problem}. 

A possible approach to alleviate the cosmological constant problem consists in replacing the cosmological constant $\Lambda_0$ by a cosmological term $\Lambda(t)$  whose value varies over time.  This can also help to decrease the severity of the coincidence problem if, for instance, the matter dilution caused by the cosmic expansion is compensated (at least partially) by the matter produced through the dark energy decay. A cosmological model with a dynamical $\Lambda$ has been proposed as early as 1933 \cite{1933} and the idea regained attention in the 1980s \cite{Hiscock,FARM,Oz,Gasp,RatPee} following the publication of studies \cite{deSit0,deSit1,deSit2,deSit3,deSit4,deSit5} suggesting that the dynamical effects of quantum fields in de Sitter space-time would lead to a decay of the effective value of the cosmological constant (see refs.~\cite{deSitU0,deSitU1,deSitU2} for an update). In the absence of a complete theory of quantum vacuum in curved spacetime, several decay laws, mostly based on empirical arguments, have been proposed in the literature.  A  review of the first $\Lambda(t)$-models developed may be found in ref.~\cite{Overduin}  and some more recent examples in refs.~\cite{n1_1,n1_2,n1_3,MMP,ZPC,selsim1,selsim2,selsim3,selsim4,XuLu,TongNoh,LuXuWu,DPZ,CDPA}.

In this study, we will consider  the phenomenological decay law
\begin{equation}
  \dot{\rho}_{\Lambda}=-Q\rho_{\Lambda}^{n}
  \label{eq:powlaw}
\end{equation}
 where $Q$ is a parameter which can be constrained from observations and $n$ an index characterizing different models. This law has been proposed in ref.~\cite{Hiscock} to study the decay of dark energy into radiation, but  it was shown that the constraints on this process are too tight to allow ${\rho_{\Lambda}}$ to decay from an initial large value to that observed today. The possibility that dark energy decays into ordinary matter is ruled out because the annihilation of matter and anti-matter would produce a $\gamma$-ray background in excess of observed level \cite{FARM}. In this paper, we will then only consider the interaction between dark energy and dark matter. Although, only the decay of dark energy into dark matter ($Q>0$) is relevant to possibly explain the  cosmological constant problem and the coincidence problem, we will also consider the decay of dark matter into dark energy ($Q<0)$ for the sake of completeness.

The model with $n=1$ has already been studied as a peculiar case of two more general models, one where the decay depends also on the cold dark matter density ($\dot{\rho}_{\Lambda}=Q_{\Lambda}\rho_{\Lambda}+Q_{c}\rho_{c}$) \cite{n1_1,n1_2} and one where the equation of state parameter of dark energy is not fixed to $-1$  ($\dot{\rho}_{x}=-Q\rho_{x}$) \cite{n1_3}. Another generalization, involving this time a generic value of $n$, has been proposed in ref.~\cite{MMP} ($\dot{\rho}_{\Lambda}=-Q\rho_{\Lambda}^{n_{\Lambda}}\rho_{m}^{n_m}$); however  the case where eq.~(\ref{eq:powlaw}) is recovered ($n_m=0$) did not have been considered. Although all these generalizations would be interesting to incorporate in our study, the resulting analysis would be significantly complicated. We will thus restrict our attention in this paper to the simple case of eq.~(\ref{eq:powlaw}).  

The model with $n=3/2$ is  particularly interesting with regard to the coincidence problem  since it is the only one where the ratio $\rho_{m}/\rho_{\Lambda}$ is constant at late time. For this reason, we will give it with a particular attention in the remainder of this article. It is to be noticed that solutions found in refs.~\cite{ZPC,selsim1,selsim2,selsim3,selsim4} where the ratio $\rho_{m}/\rho_{\Lambda}$ is constant for all time and where $w_x=w_{\Lambda}=-1$ actually correspond  to a subclass of solutions of this model  (see section~\ref{sec:MDL} for more details).

  \section{Model}
\label{sect:model}

We consider a model where dark energy and dark matter interact with an energy transfer rate $\tilde{Q}_{\Lambda}=-\tilde{Q}_{c}=-Q{\rho}_{\Lambda}^{n}$ in a spatially flat Friedmann-Robertson-Walker (FRW) spacetime. The sign of the constant $Q$ sets the direction of the transfer.  The interaction is a decay from dark energy to dark matter ($\Lambda\rightarrow$ DM) for $Q>0$ and a decay from dark matter to dark energy (DM$\rightarrow \Lambda$) for $Q<0$. The energy conservation equations for radiation ($w_r=1/3$), total matter (dark and baryonic, $w_m=0$)  and dark energy ($w_{\Lambda}=-1$)  read respectively
\begin{align}
  \dot{\rho}_{r}         &=-4H\rho_{r},        \label{eq:conrad}\\
  \dot{\rho}_{m}         &=-3H\rho_{m}+Q\rho_{\Lambda}^{n},  \label{eq:conmat}\\
  \dot{\rho}_{\Lambda}   &=-Q\rho_{\Lambda}^{n},  \label{eq:convac}
\end{align}
and the Friedmann equation takes the usual form
\begin{equation}
  \dfrac{3H^{2}}{8\pi G}=\rho_{r}+\rho_{m}+{\rho}_{\Lambda}.
  \label{eq:fried}
\end{equation}

If we set $t_0=0$ (the variables with the subscript 0 refer to their present value), the solution to eq.~(\ref{eq:convac})  is given by 
\begin{equation}
  \rho_{\Lambda}(t)= \begin{cases} 
                    \rho_{\Lambda_{0}}[1-\frac{t}{t_n}]^{-\frac{1}{n-1}}, & \mbox{ if }  n\neq1  \\
                    {\rho}_{\Lambda_{0}}e^{-Q t},  & \mbox{ if }  n=1 \\
                  \end{cases} ,
  \label{eq:denvac}
\end{equation}
where  $t_n$ is defined as
\begin{equation}
  t_n[\rho_{\Lambda_0},Q] \equiv-\dfrac{\rho_{\Lambda_{0}}^{-(n-1)}}{(n-1)Q}.
  \label{eq:trho}
\end{equation}
In the limit where $Q$ approaches 0, the time $t_n$ becomes infinite and we recover the $\Lambda$CDM model ($\rho_{\Lambda}=\rho_{\Lambda_0}$). For the decay of dark energy ($Q>0$), the time $t_n$ is situated in the future ($t_n>0$)  if $n<1$ and in the past ($t_n<0$)  if $n>1$. The converse is true for the decay of dark matter ($Q<0$). Depending on its sign, $t_n$ constitutes, or a past bound, either a future bound on the validity of eq.~(\ref{eq:denvac}). Indeed, for $n>1$ the model clearly breaks down when $t$ approaches $t_n$ since  $\rho_{\Lambda}$ becomes infinite. For $n<1$, the dark energy density is  $\rho_{\Lambda}=0$ at $t=t_n$. Hence for $Q>0$, $t_n$ corresponds to the latest time when the interaction could be physically meaningful since the dark energy would have completely decayed in dark matter at this point. Similarly, for $Q<0$, $t_n$ corresponds to the earliest time that the interaction could be meaningful.

As mentioned in the introduction, the case where $n=3/2$ is the only one which leads to a constant ratio of $\rho_{m}/\rho_{\Lambda}$  at late times and could therefore provides an elegant explanation to the coincidence problem. Indeed, for $n\neq1$ and $|t/t_n| \gg  1$, we see from eq.~(\ref{eq:denvac}) that  $\rho_{\Lambda}$ becomes proportional to $t^{-\frac{1}{n-1}}$. Moreover, at late times  $\rho_{m}$ is by hypothesis proportional to $\rho_{\Lambda}$ and the radiation can be neglected in the Friedmann equation, hence the Hubble parameter $H$ becomes proportional to $t^{-\frac{1}{2n-2}}$. It is easy to verify from eq.~(\ref{eq:conmat}) that the only value of $n$ consistent with these expressions is $3/2$ (otherwise, the term $H\rho_{m}$ would not have the same time-dependence as the other two). Because of this remarkable feature, we will consider in details the model with $n=3/2$ and find analytical solutions in order to have a deeper understanding of its characteristics. For that, we will divide the cosmological evolution into two eras: an early one dominated by radiation (RD-era), and a late one dominated by matter and dark energy (M$\Lambda$D-era). 

For all values of $n$ (including $n=3/2)$, we will use numerical solutions to constrain the models parameters. In that case, it will be more convenient  to express the energy conservation equations (\ref{eq:conrad}-\ref{eq:convac}) as functions of the redshift (corresponding to data)
\begin{align}
  \dfrac{d\rho_{r}}{dz}=&4\dfrac{\rho_r}{(1+z)},                                     \label{eq:conradz}\\
  \dfrac{d\rho_{m}}{dz}=&3\dfrac{\rho_m}{(1+z)}-Q\dfrac{\rho_{\Lambda}^{n}}{H(1+z)}, \label{eq:commatz}\\
  \dfrac{d\rho_{\Lambda}}{dz}=&Q\dfrac{\rho_{\Lambda}^{n}}{H(1+z)}.                  \label{eq:convacz}
\end{align}

\subsection{Matter$-$Dark energy dominated era (M$\Lambda$D-era): $n=3/2$}
\label{sec:MDL}

\begin{figure}[t]
\begin{minipage}[t]{0.5\linewidth}
\centering
\includegraphics[scale=0.60]{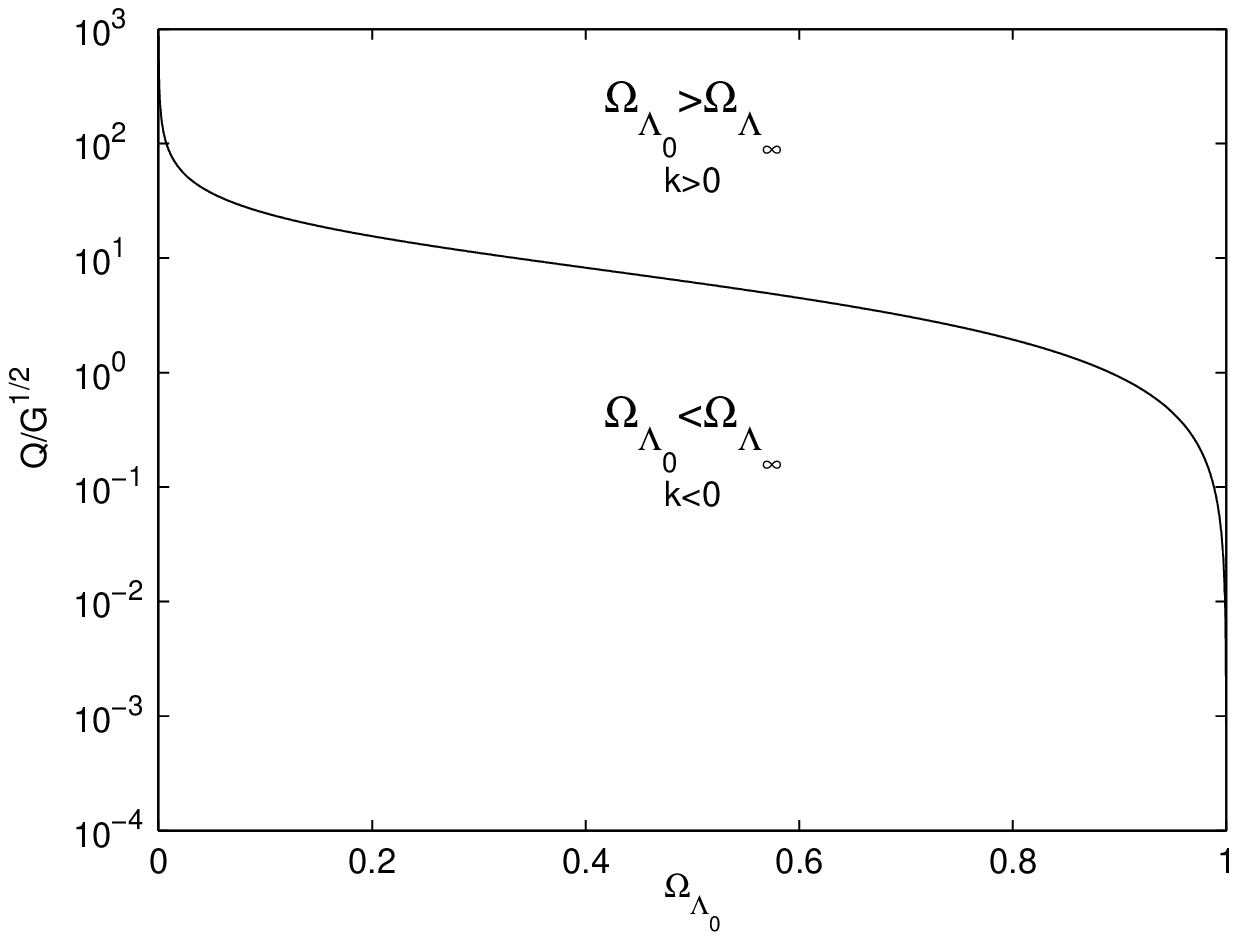}
  \caption{The curve $\Omega_{\Lambda_0}=$~$\Omega_{\Lambda_{\infty}}$ shown in the plane $\Omega_{\Lambda_0}-Q$. The parameter $Q$ is expressed here in term of the  dimensionless quantity $Q/G^{\frac{1}{2}}$, where $G$ is the Newton's constant. The physical solutions ($k<0$) are situated below the curve $\Omega_{\Lambda_0}=$~$\Omega_{\Lambda_{\infty}}$. }
  \label{fig:ksign}
\end{minipage}
\hspace{0.5cm}
\begin{minipage}[t]{0.5\linewidth}
\centering
\includegraphics[scale=0.60]{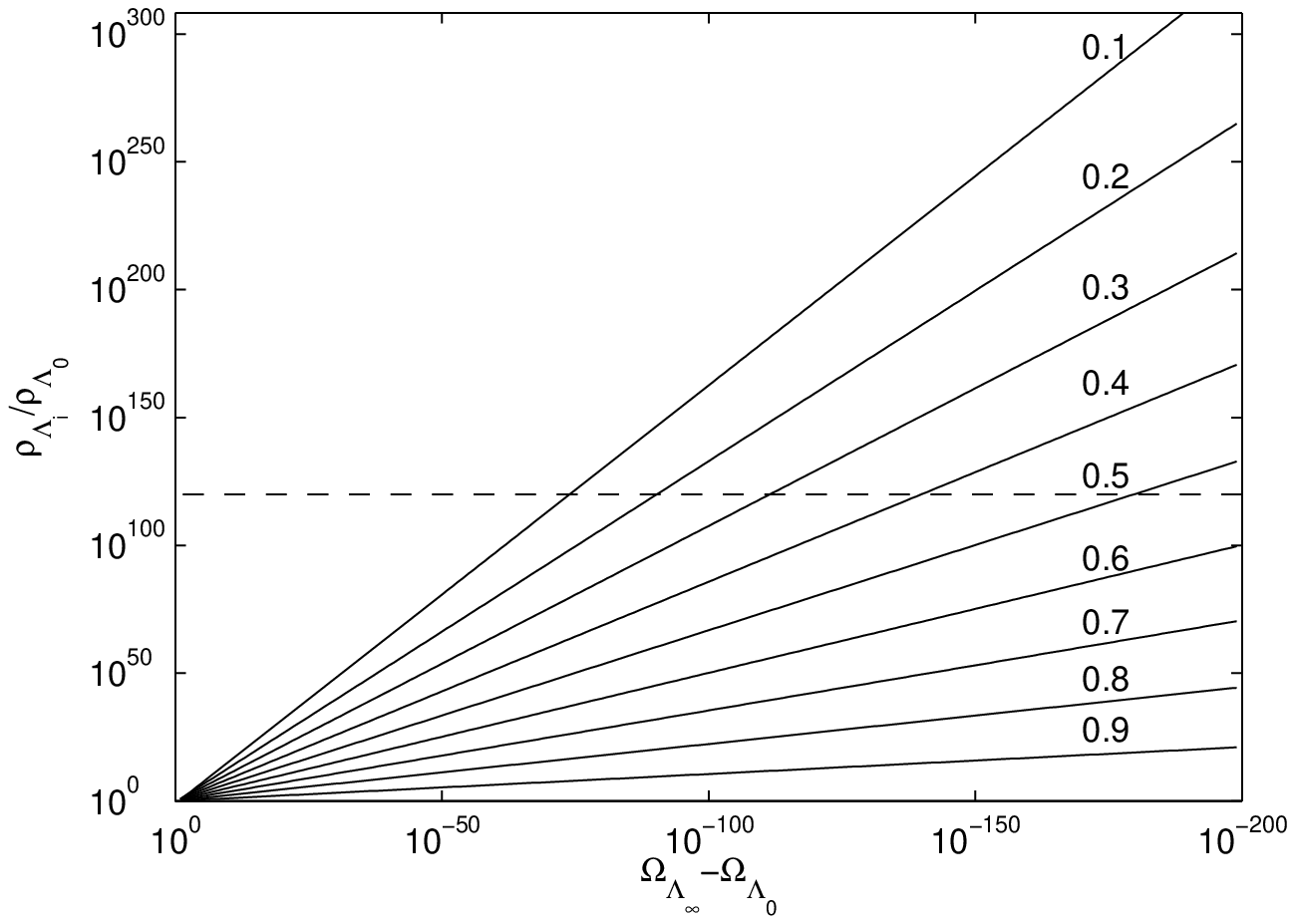}
  \caption{The ratio $\rho_{\Lambda_i}/\rho_{\Lambda_0}$ obtained from eq.~(\ref{eq:ratio}) and presented for different values of $\Omega_{\Lambda_0}$ (indicated on the figure) as a function of the difference between $\Omega_{\Lambda_{\infty}}$ and $\Omega_{\Lambda_{0}}$. The dashed line represents the ratio  between the theoretical value and the observed value of dark energy density obtained in the context of the usual $\Lambda$CDM model ($\sim10^{120}$).
}
  \label{fig:ratio}
\end{minipage}
\end{figure}

\begin{figure}[t]
  \centering
  \includegraphics[scale=0.60]{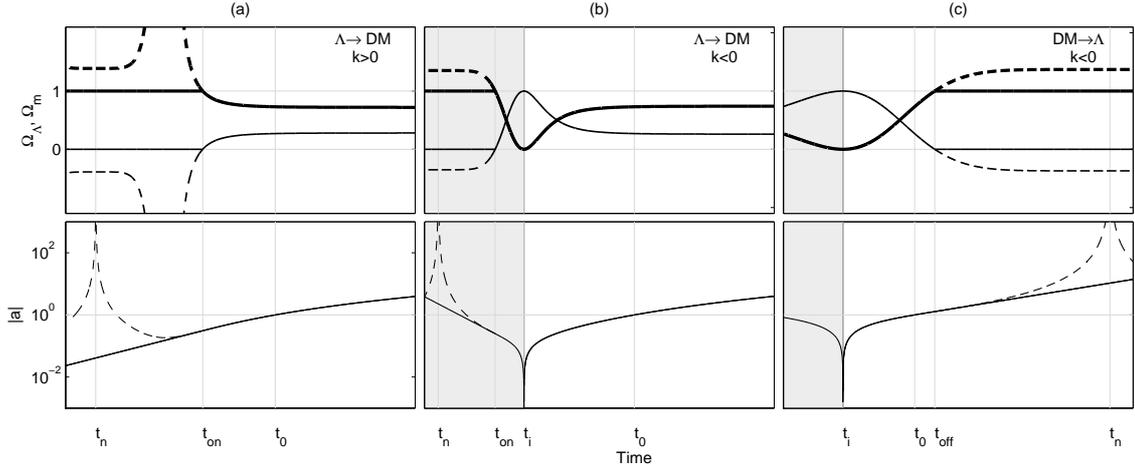}
  \caption{Examples of the evolution of the dark energy density parameter $\Omega_{\Lambda}$ (thick line), the matter density parameter $\Omega_{m}$ (thin line) and the modulus of the scale factor ($|a|$ -- for $t<t_i$, $a\in\mathbb{I}$) in absence of radiation and baryons for an expanding Universe ($H_0>0$) with (a) $\Omega_{\Lambda_{0}}>\Omega_{\Lambda_{\infty}}$, (b) $\Omega_{\Lambda_{0}}<\Omega_{\Lambda_{\infty}}<1$  and (c) $\Omega_{\Lambda_{\infty}}>1$. The dashed lines correspond to the solutions that we would obtain by letting $\rho_m$ become negative and the shaded areas to the mathematical solutions for times prior to the singularity.   We have considered that the interaction starts at $t_{\rm on}=t_w$ for (a) and stops at $t_{\rm off}=t_w$ for (c). For (a), we can notice the absence of an  initial singularity and of an early phase dominated by matter }
  \label{fig:3cases}
\end{figure}

During the M$\Lambda$D-era, we assume that $\rho_r \ll \rho_m, \rho_{\Lambda}$. We can then write the Friedman equation as
\begin{equation}
  \rho_{m}=\dfrac{3H^{2}}{8\pi G}-\rho_{\Lambda}.
  \label{eq:denmatMD}
\end{equation}
In order to have a solution for $\rho_{m}$, we must solve for $H$ first. Using the preceding equation to replace $\rho_m$ and $\dot{\rho}_m$ in eq.~(\ref{eq:conmat}) leads to the equation
\begin{equation}
  \dfrac{1}{4\pi G}\dot{H}+\dfrac{3}{8\pi G}H^{2}=\rho_{\Lambda}.
  \label{eq:diffEQ}
\end{equation}
The solution for $n=3/2$ is given by
\begin{equation}
\begin{split}
 H(t)=S_Q\left[\dfrac{\rho_{\Lambda}}{\left( \frac{3}{8\pi G}\right) }\right]^{\frac{1}{2}} \left[(\beta^{2}-1)^{-\frac{1}{2}} \left( 1+\beta \dfrac{k(\rho_{\Lambda}/\rho_{\Lambda_0})^{-\frac{\beta}{2}}-1 }{k(\rho_{\Lambda}/\rho_{\Lambda_0})^{-\frac{\beta}{2}}+1}  \right) \right]
  \end{split},
  \label{eq:HMD}
\end{equation}
where  $k$ is a dimensionless integration constant, $S_{Q}\equiv\text{sign }Q$ and $\beta\equiv S_Q\sqrt{1+96\pi G/Q^2}$. The scale factor $a$ can be obtained by integrating $H$,
\begin{equation}
 a=\tilde{a}_{0}\text{exp}\left(\int Hdt\right)=\tilde{a}_{0}\left[\rho_{\Lambda}^{\frac{-1+\beta}{6}}\left( 1+k(\rho_{\Lambda}/\rho_{\Lambda_{0}})^{-\frac{\beta}{2}}\right) ^{2/3}\right]^{S_Q}.
  \label{eq:sfac}
\end{equation}
In order to have $a=1$ at $t=t_{0}$,  the value of the constant $\tilde{a}_{0}$ is set to
\begin{equation}
 \tilde{a}_{0}=
           \left[ 
             \rho_{\Lambda_{0}}^{\frac{-1+\beta}{6}}\left( 1+k\right) ^{2/3}
           \right]^{-S_Q}.
  \label{eq:sfac0}
\end{equation}

The solutions obtained can be divided into two classes depending on the sign of the constant $k$. If we only consider dark energy and dark matter, it can be shown that the solutions with $k>0$ are characterized by the absence of an initial singularity  and by the absence of an early phase dominated by matter (which is needed for structures formation). The converse is true for $k<0$, i.e. the solutions  are characterized by the presence of an initial singularity and by the presence of an early phase dominated by matter.

To find an expression for $k$, the initial conditions $[\rho_{\Lambda_0},Q,H_0]$ could be used, however, it will be more convenient to replace the first two parameters by the current and the late-time value (supposed to be constant for $n=3/2$) of the dark energy density parameter $\Omega_{\Lambda}$. We can rearrange eq.~(\ref{eq:HMD}) in order to have
\begin{equation}
  \Omega_{\Lambda}(t)\equiv\dfrac{\rho_{\Lambda}}{\rho_{tot}}=\dfrac{\rho_{\Lambda}}{\left( \frac{3H^2}{8\pi G}\right) }=\dfrac{\beta^{2}-1}{\left[ 1+\beta\dfrac{k(\rho_{\Lambda}/\rho_{\Lambda_0})^{-\frac{\beta}{2}}-1}{k(\rho_{\Lambda}/\rho_{\Lambda_0})^{-\frac{\beta}{2}}+1}\right]^{2} }.\label{eq:xvac}  
\end{equation}
 In the limit where $t\rightarrow\infty$, $\rho_{\Lambda}$ becomes $0$ as expected and $\Omega_{\Lambda}$  approaches a constant value, $\Omega_{\Lambda}\rightarrow\Omega_{\Lambda_{\infty}}^{S_Q}$. The parameter $\Omega_{\Lambda_{\infty}}$ depends only on the value of $Q$ and is defined as
\begin{equation}
  \Omega_{\Lambda_{\infty}}\equiv\dfrac{\beta-1}{\beta+1}.
  \label{eq:xvacLT}
\end{equation}
We have to keep in mind that $\Omega_{\Lambda_{\infty}}^{S_Q}$ truly represents the late time value of $\Omega_{\Lambda}(t)$ only for the decay of dark energy. For the decay of dark matter, we have shown  that the model breaks down in the future at $t=t_{n}$, hence before reaching the ``late time" (actually, as we will see, the model breaks down even before $t_{n}$). In this case, $\Omega_{\Lambda_{\infty}}$ has to be considered only as a parameter related to the strength of the interaction.
Setting $t=t_0$ in eq.~(\ref{eq:xvac}) and solving for $k$  yields
\begin{equation}
  k[\Omega_{\Lambda_{0}},\Omega_{\Lambda_{\infty}},H_0]= \dfrac{ S_{H_0}\Omega_{\Lambda_{\infty}}\Omega_{\Lambda_{0}}^{\frac{1}{2}}+\Omega_{\Lambda_{\infty}}^{\frac{1}{2}}}
{S_{H_0}\Omega_{\Lambda_{0}}^{\frac{1}{2}}-\Omega_{\Lambda_{\infty}}^{\frac{1}{2}}},
  \label{eq:komegainf}
\end{equation}
where $S_{H_0}\equiv\text{sign }H_0$. For an expanding Universe ($H_0>1$), the sign of $k$ depends only on whether $\Omega_{\Lambda_{\infty}}$ is greater or smaller than $\Omega_{\Lambda_{0}}$. Since  $\Omega_{\Lambda_{0}}$  and $\Omega_{\Lambda_{\infty}}^{S_Q}$ are both included in the interval $[0,1]$, $k$ is necessarily negative for the decay of dark matter but could  be either positive (if $\Omega_{\Lambda_{0}}>\Omega_{\Lambda_{\infty}}$) or negative (if $\Omega_{\Lambda_{0}}<\Omega_{\Lambda_{\infty}}$) for the decay of dark energy. The solutions with $k=0$ correspond to a situation where the ratio $\rho_m/\rho_{\Lambda}$ is constant for all time (at least as long as radiation is neglected) and which is a peculiar case of the solutions found in refs.~\cite{ZPC,selsim1,selsim2,selsim3,selsim4}. The curve $\Omega_{\Lambda_0}=$~$\Omega_{\Lambda_{\infty}}$ is shown in the plane $\Omega_{\Lambda_0}-Q$ in figure~\ref{fig:ksign}. For the decay of dark energy, the relevant range of values for $Q$ is roughly situated between $10^{-2}G^{\frac{1}{2}}$ and $10^{2}G^{\frac{1}{2}}$. Below the lower bound, the interaction becomes insignificant and we recover approximately the $\Lambda$CDM model ($\Omega_{\Lambda_{\infty}}\approx1$), whereas above the upper bound, $\Omega_{\Lambda}$ remains close to zero for all time.

From eq.~(\ref{eq:sfac}) it should be obvious that only the solutions with $k<0$ feature a singularity and that it occurs at the time
\begin{equation}
  t_i=\left[1-\left(-\dfrac{1}{k}\right)^{\frac{1}{\beta}}\right]t_{n}.
  \label{eq:ti}
\end{equation}
For $H_{0}>0$, this singularity occurs in the past ($t_i<t_0$) and could be considered as the initial time of the Universe. Evaluating eq.~(\ref{eq:denvac}) at $t=t_i$, we  find an expression for the ratio of the initial and the current value of the dark energy density 
\begin{equation}
  \dfrac{\rho_{\Lambda_i}}{\rho_{\Lambda_0}}=\left(-k \right)^{\frac{2}{\beta}}= \left[ \dfrac{\Omega_{\Lambda_{\infty}}\Omega_{\Lambda_{0}}^{\frac{1}{2}}+\Omega_{\Lambda_{\infty}}^{\frac{1}{2}}}{\Omega_{\Lambda_{\infty}}^{\frac{1}{2}}-\Omega_{\Lambda_{0}}^{\frac{1}{2}}} \right]^{\frac{2}{\beta}}  .
  \label{eq:ratio}
\end{equation}

The case where $\Omega_{\Lambda_0}\approx\Omega_{\Lambda_{\infty}}$ could in principle provide an explanation for both the coincidence and the cosmological constant problems. Indeed, when radiation is neglected, we have $\rho_{m}/\rho_{\Lambda}=1/\Omega_{\Lambda}-1$, hence the condition $\Omega_{\Lambda_0}\approx\Omega_{\Lambda_{\infty}}$ implies that the current value of the ratio   $\rho_{m}/\rho_{\Lambda}$ is now typical of a large period of time and there is no more ``coincidence'' problem. Moreover, if $\Omega_{\Lambda_0}\approx\Omega_{\Lambda_{\infty}}$, the ratio $\rho_{\Lambda_i}/\rho_{\Lambda_0}$ can be arbitrarily large. However, if the difference between the theoretical value of $\rho_{\Lambda_i}$ and the value of $\rho_{\Lambda_0}$ obtained for the $\Lambda(t)$CDM model remains of several orders of magnitude, we see from figure~\ref{fig:ratio} that a fine tuning of the quantity $\Omega_{\Lambda_{\infty}}-\Omega_{\Lambda_0}$ would be needed to explain it.

To complete the analysis of the M$\Lambda$D-era, we must come back to the problem encountered in the future for the decay of dark matter. It easy to verify that at $t_{n}$, the density parameter of dark energy is greater than 1, which constitutes a violation of the null energy condition ($w<-1)$. Setting $\Omega_{\Lambda}=1$ in eq.~(\ref{eq:xvac}) and solving for $t$, we actually find that the null energy condition starts to be violated at
\begin{equation}
 t_w=\left[1-\left(-\dfrac{k_w}{k}\right)^{\frac{1}{\beta}}\right]t_{n},
  \label{eq:tw}
\end{equation}
where
\begin{equation}
 k_w[\Omega_{\Lambda_{0}},\Omega_{\Lambda_{\infty}},H_0]\equiv\dfrac{S_k S_Q\Omega_{\Lambda_{\infty}}^{\frac{1}{2}} + \Omega_{\Lambda_{\infty}} }{S_k S_Q\Omega_{\Lambda_{\infty}}^{\frac{1}{2}} -1}
  \label{eq:kw}
\end{equation}
and $S_k$ is defined as the sign of the constant $k$. The density parameter $\Omega_{\Lambda}$ could become greater than 1 only if we let $\rho_m$   become negative after $t_w$. To prevent this problem, when the dark matter has completely decayed in dark energy ($\rho_{c}=0$), we should consider that there is no more interaction and the dark energy density becomes a constant. If we neglect the baryon density, the interaction stops at $t_{\rm off}=t_w$ and the late time($t>t_{\rm off}$)  value of $\rho_{\Lambda}$ is then given by
\begin{equation}
  \rho_{\Lambda_{w}}=\left(-\dfrac{k}{k_w}\right)^{\frac{2}{\beta}}\rho_{\Lambda_0}.
  \label{eq:rhow}
\end{equation}

The solutions for $\Omega_{\Lambda}$, $\Omega_{m}=1-\Omega_{\Lambda}$  and $|a|$ are shown in figure \ref{fig:3cases} for  $\Omega_{\Lambda_{0}}>\Omega_{\Lambda_{\infty}}$, $\Omega_{\Lambda_{0}}<\Omega_{\Lambda_{\infty}}<1$ and $\Omega_{\Lambda_{\infty}}>1$. As we can see, the null energy condition is also violated for the decay of dark energy, but this time the violation occurs in the past. The presence of the singularity prevents this problem ($t_i>t_w$) for $k<0$ but not for $k>0$.   We could simply consider that $\rho_{\Lambda}$ is constant before $t_{\rm on}=t_w$. However, the reason why the dark energy should start to decay at $t_w$ would be unclear. It would also be possible to consider that decay starts after $t_w$ (when $\rho_m$ is non-zero), which would lead to a solution with an initial singularity and an early phase dominated by matter. That could be the case for instance if, for some reason, the dark energy decay starts only when the dark matter becomes sufficiently diluted. However,  we do not have considered any threshold of this kind for the case $k<0$. In order to be consistent and to keep our model as simple as possible, we will not introduce any threshold, but rather consider that the model is not valid for $k>0$.

\subsection{Radiation dominated era (RD-era): $n=3/2$}

\begin{figure}[t]
  \centering
  \includegraphics[scale=0.55]{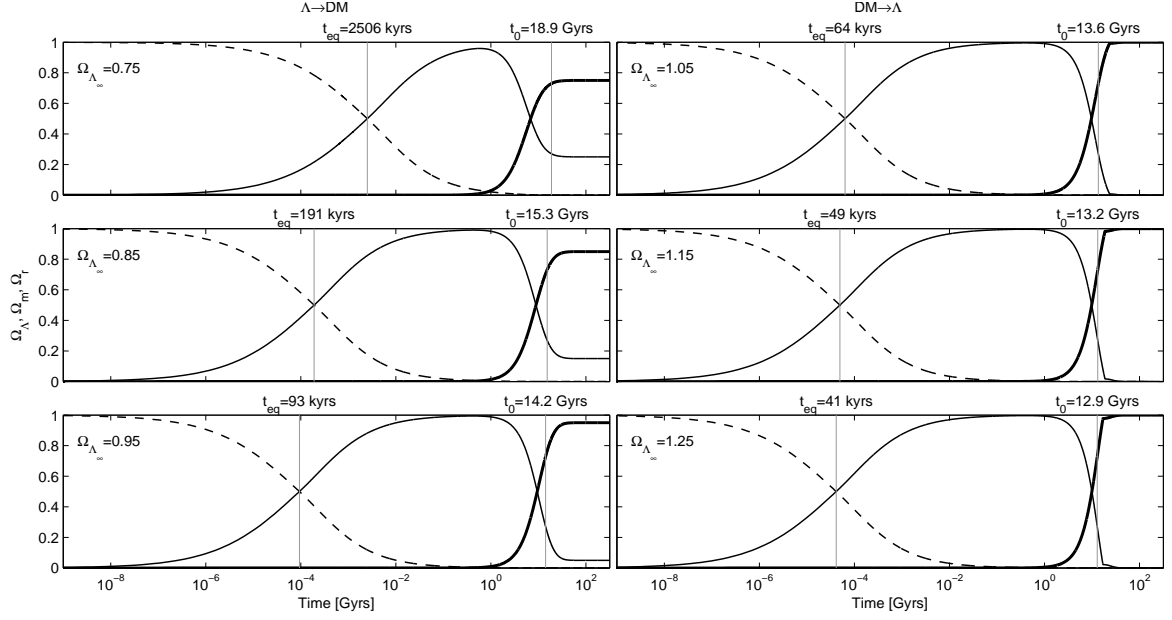}
  \caption{Evolution of dark energy ($\Omega_{\Lambda}$, thick line), matter ($\Omega_{m}$, thine line) and radiation ($\Omega_{r}$, dashed line) density parameter  for different values of  $\Omega_{\Lambda_{\infty}}$ and with $\Omega_{\Lambda_{0}}=0.73$ and $H_0=70\text{ km}\ \text{s}^{-1} \text{Mpc}^{-1}$.  The values of $\Omega_{b_{0}}$ and $\Omega_{r_{0}}$ are set using eq.~(\ref{eq:Ob0}) and eq.~(\ref{eq:Or0}). The time has been redefined setting $t_i=0$, hence the age of the Universe is given by $t_0$.}
  \label{fig:6RML}
\end{figure}

Until now, the radiation has been neglected. This approximation clearly does not hold at early times. Solving eq.~(\ref{eq:conrad}) leads to the  usual result (which holds for all time)
\begin{equation}
  \rho_{r}=\rho_{r_{0}}a^{-4}.
  \label{eq:denrad}
\end{equation}
During the RD-era, we can neglect the dark energy and the matter densities ($\rho_r \gg \rho_m, \rho_{\Lambda}$). Using the Friedmann equation, we can easily find the Hubble parameter 
\begin{equation}
  H=2H_{r_0}a^{-2}.
  \label{eq:HRD}
\end{equation}
where $H_{r_0}\equiv\sqrt{(8\pi G/3)\rho_{r_0}}$ and integrate it to obtain an expression for the scale factor
\begin{equation}
  a=\left[2H_{r_0}(t-t_{eq}) +a_{eq}^{2}  \right]^{\frac{1}{2}}.
\end{equation} 
The value of the integration constant has been set in order to have $a_{eq}\equiv a_{M\Lambda D}(t_{eq})=a_{RD}(t_{eq})$, where $t_{eq}$ is defined as the time where $\rho_r=\rho_{\Lambda}+\rho_m$. The initial singularity now occurs at 
\begin{equation}
   t_{i}=t_{eq}-\dfrac{a^{2}_{eq}}{2H_{r_0}}.
\end{equation}
We cannot find an analytical expression for $t_{eq}$; instead, we have to make the approximation that the solutions found for $a$ and $H$ during the M$\Lambda$D-era hold up to $t_{eq}$  and solve numerically the equation
\begin{equation}
\begin{split}
   \rho_{r_{M\Lambda D}}(t_{eq})&=\rho_{m_{M\Lambda D}}(t_{eq})+\rho_{\Lambda_{M\Lambda D}}(t_{eq}),\\
   \rho_{r_{0}}a_{M\Lambda D}^{-4}(t_{eq})&=\dfrac{3H^{2}_{M\Lambda D}(t_{eq})}{8\pi G}.
   \label{eq:transc}
\end{split}
\end{equation}
To find an expression for the matter density $\rho_{m}$ during the RD-era,  we can use the expressions found for $\rho_{\Lambda}$ and for $H$ (eqs.~(\ref{eq:denvac}) and (\ref{eq:HRD})) to solve the matter conservation equation (eq.~(\ref{eq:conmat})). Setting the value of the integration constant in order to have $\rho_{m_{eq}}\equiv\rho_{m_{RD}}(t_{eq})=\rho_{m_{M\Lambda D}}(t_{eq})$, we get
\begin{equation}
    \rho_{m}=\rho_{m_{eq}}\left( \dfrac{a}{a_{eq}}\right) ^{-3} + \left[F_1(t)+F_2(t) \right] a^{-3}, 
\label{eq:rm_RD}
\end{equation}
where $F_1$ is given by
\begin{equation}
F_1\equiv\left(\rho_{\Lambda_{eq}}a_{eq}^3-\rho_{\Lambda}a^3 \right)+3H_{r_0}|t_{n}|\rho_{\Lambda_0}^{\frac{1}{2}} \left(\rho_{\Lambda_{eq}}^{\frac{1}{2}}a_{eq}-\rho_{\Lambda}^{\frac{1}{2}}a \right)
\end{equation}
and $F_2$ by
\begin{equation}
  F_2\equiv\left( \dfrac{6H_{r_0}^2t_{n}^2}{\alpha}\right) \left[\text{arctan}\left(\dfrac{a}{\alpha} \right) - \text{arctan}\left(\dfrac{a_{eq}}{\alpha} \right) \right]\rho_{\Lambda_0},\quad
\text{with }\alpha\equiv\sqrt{\left|2H_{r_0}(t_{eq}-t_{n})-a_{eq}^{2}\right|}.
\end{equation}
\nolinebreak In the limit where $Q$ approaches 0, $F_1$ and $F_2$ are reduced to $F_1=-F_2=3H_{r_0}|t_{n}|(a_{eq}-a)\rho_{\Lambda_0}$, so the second term in eq.~(\ref{eq:rm_RD}) vanishes and we recover the expected expression for the  $\Lambda$CDM model.

As a consequence of the approximations made for the M$\Lambda$D-era and for the RD-era, $\rho_{tot}\neq3H^{2}/(8\pi G)$ in the vicinity of $t_{eq}$. To evaluate $\rho_{tot}$ (and $\Omega_i$) we should instead explicitly add $\rho_{\Lambda}$, $\rho_{m}$ and $\rho_{r}$ together. The evolution of the density parameters are shown in figure \ref{fig:6RML} for different values of $\Omega_{\Lambda_{\infty}}$. For smaller values of $\Omega_{\Lambda_{\infty}}$, the age of the Universe is larger. There is also an inverse relationship relating $\Omega_{\Lambda_{\infty}}$ and the duration of the RD-era and of the $\Lambda$D-era, whereas the duration of the MD-era decreases as $\Omega_{\Lambda_{\infty}}$ becomes smaller. The variation of these time intervals becomes increasingly important as  $\Omega_{\Lambda_{\infty}}$ approaches $\Omega_{\Lambda_{0}}$, leading to solutions significantly different from what we would have obtained in the $\Lambda$CDM model, even for $t<t_0$. We can then expect that these solutions  will be disfavoured by the observational constraints.

  \section{Observational Constraints}

In this section, we explain our methodology for using the currently available data to constrain the three free parameters ($\Omega_{\Lambda_0}$, $\Omega_{\Lambda_{\infty}}$, $H_0$) of the model. Following what  was done in refs.~\cite{DPZ,XuLu,TongNoh,LuXuWu} to constrain cosmological models with interacting dark sectors, we will consider the tests described below involving the distance modulus $\mu$ of type Ia supernova (SNeIa) and  gamma-ray bursts (GRB), the baryon acoustic oscillation (BAO), the cosmic microwave background (CMB) and the observational Hubble rate (OHD). In refs.~\cite{DPZ,XuLu,TongNoh,LuXuWu}, the authors also use the gas mass fractions in galaxy clusters as inferred from x-ray data to constrain their models, however since this method involves many uncertain parameters, we do not consider it here. The best fit follows from minimizing the sum of the $\chi^2$ of each data set
\begin{equation}
   \chi^{2}_{tot}=\chi^{2}_{\mu}+\chi^{2}_{OHD}+\chi^{2}_{BAO}+\chi^{2}_{CMB}.
   \label{eq:Xtot}
\end{equation}

\subsection{Distance modulus $\mu$ of SNeIa and GRB}
The distance modulus is the difference between the apparent magnitude $m$ and the absolute magnitude $M$ of an astronomical object. Its theoretical value for a flat Universe is given by
\begin{equation}
  \mu_{th}(z)\equiv5\log_{10}\dfrac{D_{L}(z)}{h}+42.38,
  \label{eq:mut}
\end{equation}
where $h=H_{0}$/(100 $\text{km}\ \text{s}^{-1} \text{Mpc}^{-1}$) and the Hubble free luminosity distance $ D_{L}$ is defined as
\begin{equation}
  D_{L}\equiv H_{0}(1+z)\int^{z}_{0}\dfrac{dz}{H}.
  \label{eq:DL}
\end{equation}
The best fit is obtained by minimizing the $\chi^{2}$ function
\begin{equation}
  \chi^{2}_{\mu}[\Omega_{\Lambda_{0}},\Omega_{\Lambda_{\infty}},H_{0}]=\sum\limits_{i} \dfrac{[\mu_{obs}(z_{i})-\mu_{th}(z_{i})]^{2}}{\sigma_{i}^{2}},
  \label{eq:ChiMu}
\end{equation}
where $\sigma_{i}$ is the  1-$\sigma$ uncertainty associated with $i^{th}$ point. The observational data used are the 557 distance modulii of SNeIa assembled in the Union2 compilation \cite{SNE5} ($0.015 < z < 1.40$) and the 59 distance modulii of GRB from \cite{GRB} ($1.44 < z < 8.10$).  The combination of these two types of data covers a wide range of redshift  providing a more complete description of the cosmic evolution than the SNeIa data by themselves.

\subsection{Observational $H(z)$ data (OHD)}

\begin{table}[t]
\begin{footnotesize}
\begin{center}
\begin{tabular}{c | c c c c c c c c c c c c c c c}
\toprule
    $z$ 
        & $\ \ \quad 0$ \cite{Reiss} & 0.1     & 0.17     & $\ \ \quad 0.24$ \cite{Gaztanaga}         & 0.27      & 
          $\ \ \quad 0.34$ \cite{Gaztanaga}       & 0.4       &  $\ \ \quad 0.43$ \cite{Gaztanaga}     \\ 
    $H(z)$ ($\text{km}\ \text{s}^{-1} \text{Mpc}^{-1}$)                
        & 74.2         & 69        & 83       & 79.69      & 77        & 83.80       & 95        &  86.45    \\ 
    1-$\sigma$ uncertainty 
        & $\pm$3.6     & $\pm$12   & $\pm$8   & $\pm$3.61  & $\pm$14   & $\pm$4.55   & $\pm$17   & $\pm$4.96 \\
\midrule
    $z$                     
        & 0.48         & 0.88      & 0.9       & 1.3       & 1.43      & 1.53      & 1.75    \\ 
    $H(z)$ ($\text{km}\ \text{s}^{-1} \text{Mpc}^{-1}$)               
        & 97           &  90        & 117       & 168       & 177       & 140       & 202     \\ 
    1-$\sigma$ uncertainty   
        & $\pm$62      &  $\pm$40   & $\pm$23   & $\pm$17   & $\pm$18   & $\pm$14   & $\pm$40 \\
\bottomrule
\end{tabular}
\end{center}
\end{footnotesize}
\caption{Observational $H(z)$ data used to constrain our model. Unless indicated otherwise, the values are taken from table 2 in \cite{Stern}.}
\label{tab:OHD}
\end{table}

As a function of the redshift, the Hubble parameter is given by
\begin{equation}
  H(z)=-\dfrac{1}{1+z}\dfrac{dz}{dt}.
  \label{eq:Hz}
\end{equation}
Therefore it is  possible to  measure $H(z)$  through a determination of $dz/dt$. In \cite{Jimenez}, Jimenez et al. demonstrated the feasibility of the method by applying it to a $z\sim0$ sample. Simon et al. \cite{Simon}, completed by Stern et al. \cite{Stern} have applied this method using the differential ages of passively evolving galaxies to derive a set of 11 observational values for $H(z)$ in the range $0.1 < z < 1.8$.  In addition to these values, we also constrain the model using the value of $H$ at $z=0$ obtained by Riess et al. \cite{Reiss} from the observations of 240 Cepheids, as well those obtained  by Gazsta\~{n}aga et al. \cite{Gaztanaga} at $z=0.24$, $z=0.34$ and $z=0.43$ using the radial BAO peak scale as a standard ruler (see  ref.~\cite{Gaztanaga} for more details). The best fit is obtained by minimizing the $\chi^{2}$ function
\begin{equation}
  \chi^{2}_{OHD}[\Omega_{\Lambda_{0}},\Omega_{\Lambda_{\infty}},H_{0}]=\sum\limits_{i} \dfrac{[H_{obs}(z_{i})-H_{th}(z_{i})]^{2}}{\sigma_{i}^{2}}.
  \label{eq:ChiMu}
\end{equation}
The data used are summarized in table \ref{tab:OHD}.

\subsection{Baryon acoustic oscillation  (BAO)}
The use of BAO to test dark energy models is usually made by means of the distance parameter $\mathcal{A}$. However, as pointed out in ref.~\cite{CDPA}, $\mathcal{A}$ is not appropriate to test a model with matter production associated with the decays of dark energy. Instead, we can use the dilation scale
\begin{equation}
  D_{V}(z)=c\left[\dfrac{z}{H(z)}\left(\int_{0}^{z}\dfrac{dz}{H(z)} \right)^{2}  \right]^{1/3}.
\end{equation}
The ratio  $r_s(z_d)/D_V(z)$, where $r_s(z_d)$ is the comoving sound horizon size at the drag epoch,  has been observed at $z=0.35$ by SDSS \cite{Einsenstein} and at $z=0.20$ by 2dFGRS \cite{DV_ratio}. It would be possible to proceed as in refs.~\cite{XuLu,LuXuWu} and use a fitting formula (developed for the $\Lambda$CDM model) to find $r_s(z_d)$. However we can avoid this if we follow refs.~\cite{DPZ,TongNoh} and minimize the $\chi^2$ of the ratio $D_{V_{0.35}}/D_{V_{0.20}}$ given by
\begin{equation}
  \chi_{BAO}^{2}[\Omega_{\Lambda_{0}},\Omega_{\Lambda_{\infty}}]=\dfrac{[ (D_{V_{0.35}}/D_{V_{0.20}})_{th}-(D_{V_{0.35}}/D_{V_{0.20}})_{obs}]^2}{\sigma^{2}_{0.35/0.20}}.
\end{equation}
In addition to being independent of $r_s$, this ratio is also independent of $H_0$. The observed value for $D_{V_{0.35}}/D_{V_{0.20}}$ is $1.736\pm0.065$ \cite{DV_ratio}.

\subsection{Cosmic Microwave Background (CMB)}
\label{sec:CMB}

The values extracted from the 7-year WMAP data for the acoustic scale ($l_A(z_*)$), for the CMB shift parameter ($R$), and for the  redshift at the decoupling epoch ($z_*$) can be used to constrain the model parameters. The CMB shift parameter and the acoustic scale are respectively defined as
\begin{equation}
  R=\sqrt{\Omega_{m_{0}}H_{0}^2}\int_0^{z_*}\dfrac{dz}{H}
\end{equation}
and
\begin{equation}
  l_A=\dfrac{\pi\int^{z_*}_{0}\frac{dz}{H}}{\int^{\infty}_{z_*}\frac{c_s}{c}\frac{dz}{H}}.
\end{equation}
Since the  sound velocity $c_s$ is given by
\begin{equation}
  c_s=c\left(3+\dfrac{9}{4}\dfrac{\Omega_{b_0}}{\Omega_{\gamma_0}(1+z)}\right)^{-1/2}, 
\end{equation}
two additional free parameters are needed to determine the acoustic scale, namely the current value of the density parameter of baryons ($\Omega_{b_0}$) and of radiation ($\Omega_{\gamma_0}$). Constraining the model with these two additional parameters will require in an increased computational cost. However as suggested in ref.~\cite{CDPA}, we can use the values obtained in the context of the $\Lambda$CDM cosmology. This is motivated since the radiation and the baryons are separately conserved, and because we want to preserve the spectrum profile as well the nucleosynthesis constraints. The observational results from 7-year WMAP data \cite{CMB4} are 
\begin{equation}
  \Omega_{b_{0}}=2.25\times 10^{-2}h^{-2} \quad \text{ and } \quad \Omega_{\gamma_{0}}=2.469\times 10^{-5}h^{-2}.
\label{eq:Ob0}
\end{equation}

Considering the high redshift values involved here, the dynamical effects of radiation cannot be neglected. We therefore must include the density parameter of radiation, $\Omega_{r_{0}}$, as an extra initial parameter. However, this quantity is related to the density parameter of photons through 
\begin{equation}
  \Omega_{r_{0}}=\left( 1+\dfrac{7}{8}\left(\dfrac{4}{11}\right)^{\frac{4}{3}}N_{\rm eff}\right) \Omega_{\gamma_{0}}.
  \label{eq:Or0}
\end{equation}
Hence, providing that the effective number of neutrino species, $N_{\rm eff}$, is determined, we still have only three parameters to constrain. The departure of $N_{\rm eff}$ from 3 is due to neutrino heating by $e^{\pm}$ annihilations in early Universe.
The value inferred from observations, usually close to 3 (3.04 \cite{Mangano}, 3.14 \cite{cyburt}),  was recently updated to 4.34 \cite{CMB4}.

The decoupling epoch occurs when the expansion time becomes less than the Thomson scattering time. From a practical point of view, $z_*$ may be computed by finding the redshift value corresponding to an optical depth of unity
\begin{equation}
  \tau(z_*)=\int^{z_*}_{0}\dfrac{n_e\sigma_{T}c}{(1+z)H}dz=1,
  \label{eq:OpD}
\end{equation}
where $\sigma_{T}$ is the Thompson cross-section. To evaluate this integral, we need to know the evolution of the electron density $n_e$ as a function of the redshift, which is not simple because of its dependence on the recombination process. To avoid to do this complex computation each time that one need to know the value of $z_*$, a fitting formula has been developed for the $\Lambda$CDM model \cite{fitting}
\begin{equation}
   z_*=1048[1+0.00124(\Omega_{b_0} h^2)^{-0.738}][1+g_1(\Omega_{m_0} h^2)^{g_2}],
   \label{eq:fitz}
\end{equation}
where
\begin{alignat}{3}
g_1 & \equiv0.0783(\Omega_{b_0} h^2)^{-0.238}(1+39.5(\Omega_{b_0} h^2)^{-0.763})^{-1},\\
g_2 & \equiv0.560(1+21.1(\Omega_{b_0} h^2)^{1.81})^{-1}.
\end{alignat}
Following refs.~\cite{XuLu,LuXuWu}, we approximate the value of $z_*$ in the $\Lambda(t)$CDM  model using eq.(~\ref{eq:fitz}). However, from eq.~\ref{eq:OpD}, it is obvious that even if we set all the parameters, except $Q$, to the same values, the decoupling redshift for the interacting and the non-interacting cases will be different as a consequence of a different redshift-dependence for the Hubble parameter. Hence, the validity of this approximation may be questioned; we will come back to this issue in the results section.

Defining $v\equiv(l_A-l_A^{obs}, R-R^{obs},z_*-z_*^{obs})$, the best fit is obtained by minimizing the $\chi^{2}$ function
\begin{equation}
  \chi^{2}_{CMB}[\Omega_{\Lambda_{0}},\Omega_{\Lambda_{\infty}},H_{0}]=v M v^t,
\end{equation}
where $M$ is the inverse variance-covariance matrix from the 7-year WMAP data
\begin{equation}
M=\left(
\begin{array}{rrr}
2.305 & 29.698 & -1.333 \\
29.698 & 6825.270 & -113.180 \\
-1.333 & -113.180 & 3.414 
\end{array} 
\right) 
\end{equation}
and the observed value are also taken from the 7-year WMAP data, $l_A(z_*)=302.09\pm0.76$, $R(z_*)=1.725\pm0.018$ and $z_*=1091.3\pm0.91$.

  \section{Results and discussion}

\subsection{Case $n=3/2$}
  \label{sect:n32}

\begin{figure}[t]
  \centering
  \includegraphics[scale=0.60]{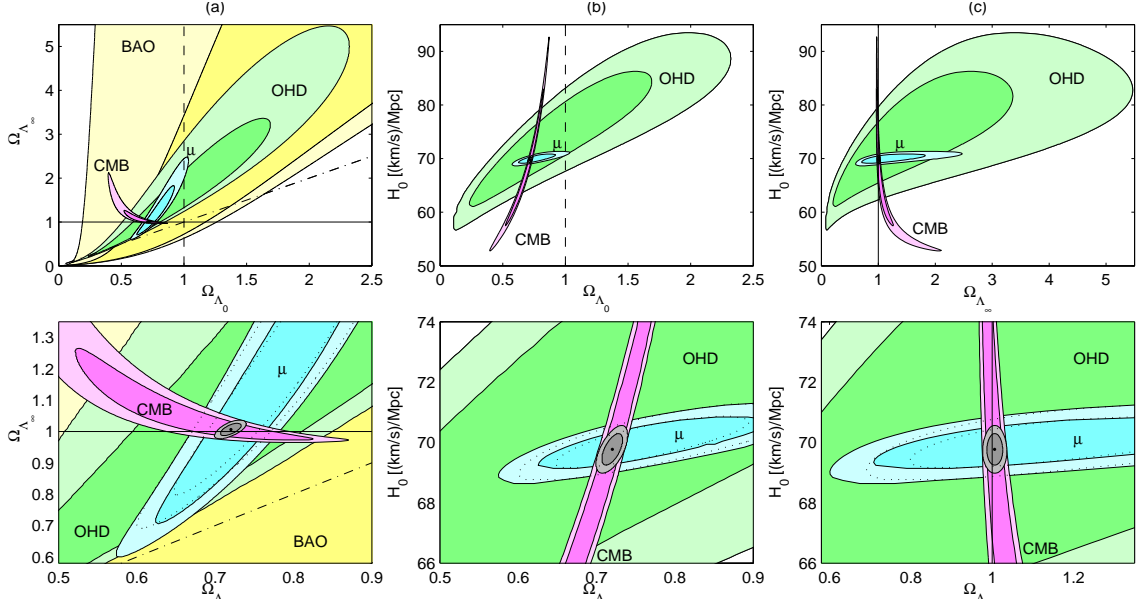}
  \caption{Projection (a) in the plan $\Omega_{\Lambda_0}-\Omega_{\Lambda_{\infty}}$, (b) in the plan $\Omega_{\Lambda_0}-H_0$ and (c) in the plan $\Omega_{\Lambda_{\infty}}-H_0$ of the 1-$\sigma$ (darker colour) and the 2-$\sigma$ (lighter colour) confidence region obtained from four different types of observational data, the distance modulus $\mu$ (blue), BAO (yellow), CMB (pink), OHD (green) and the from combination of all of them ($C_4$, $\mu$+OHD+BAO+CMB) (gray). The confidence regions obtained from combination of observational constraints $C_3$ ($\mu$+OHD+BAO) are shown in the lower panels (dotted contours).  The region situated at the right of the dashed  line ($\Omega_{\Lambda_0}=1$), involved negative matter energy density and must be considered as non-physical. The solid  line ($\Omega_{\Lambda_{\infty}}=1$) sets the separation where dark energy is decaying in dark matter (below the line) and where dark matter is decaying in dark energy (below the line). Below the dot-dashed line ($\Omega_{\Lambda_0}=\Omega_{\Lambda_{\infty}}$), $k>0$ and above,  $k<0$. The best fit parameters are also indicated by a black dot in each panel.}
\label{fig:confreg}
\end{figure}

\begin{figure}[t]
  \centering
  \includegraphics[scale=0.70]{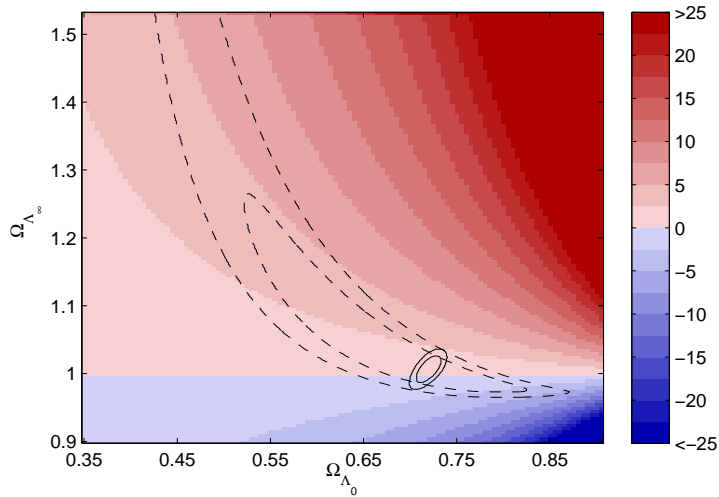}
  \caption{Percentage of difference on the value of the integral $\int^{z_*}_{0}dz/(H(z)(1+z))$ evaluated for the $\Lambda$CDM model and for the $\Lambda(t)$CDM model for a corresponding value of $\Omega_{\Lambda_0}$ ($z_*$ is given by  eq.~(\ref{eq:fitz})). A positive difference means that the $\Lambda$CDM value is larger than the $\Lambda(t)$CDM one and a negative difference, the converse. The dashed lines and the solid lines represent respectively, the 1-$\sigma$ and the 2-$\sigma$ confidence contours obtained from the CMB data, and from the combination of data $C_4$ ($\mu$+OHD+BAO+CMB).}
\label{fig:aSCALE}
\end{figure}

\begin{table}[t]
\begin{footnotesize}\begin{center}
\renewcommand
\arraystretch{1.5}
\begin{tabular}{c|ccccr}
\toprule
& $N_{\rm eff}$ & $\Omega_{\Lambda_0}$ & $\Omega_{\Lambda_{\infty}}$ & $H_0$ ($\text{km}\ \text{s}^{-1} \text{Mpc}^{-1}$) & \multicolumn{1}{c}{$\chi^2_{\rm min}$}  \\
\midrule
$C_3$  & 3.04                                                &
         0.723 $ ^{+0.035}_{-0.033}$ $ ^{+0.051}_{-0.051}$   & 
         1                                                   & 
         69.82 $ ^{ +0.61}_{ -0.50}$ $ ^{ +1.00}_{ -0.91}$   &
         576.178  \\
$C_4$  & 3.04                                                &
         0.718 $ ^{+0.009}_{-0.012}$ $ ^{+0.014}_{-0.019}$   & 
         1                                                   & 
         69.85 $ ^{ +0.49}_{ -0.55}$ $ ^{ +0.76}_{ -0.82}$   &
         576.638  \\
$C_4$  & 3.14                                                &
         0.712 $ ^{+0.009}_{-0.012}$ $ ^{+0.014}_{-0.017}$   & 
         1                                                   & 
         69.75 $ ^{ +0.48}_{ -0.57}$ $ ^{ +0.75}_{ -0.83}$   &
         577.083  \\
$C_4$  & 4.34                                                &
         0.635 $ ^{+0.012}_{-0.014}$ $ ^{+0.018}_{-0.021}$   & 
         1                                                   & 
         69.02 $ ^{ +0.51}_{ -0.55}$ $ ^{ +0.78}_{ -0.83}$   &
         613.694 \\
\bottomrule
\end{tabular}
\end{center}\end{footnotesize}
\caption{Best fit values for the $\Lambda$CDM model parameters and the corresponding $\chi_{\rm min}^{2}$ for three different values of $N_{\rm eff}$ inferred from observations 3.04 \cite{Mangano}, 3.14 \cite{cyburt}, and 4.34 \cite{CMB4}. We have considered two different combinations of observational constraint, $C_3$ ($\mu$+OHD+BAO) and $C_4$ ($\mu$+OHD+BAO+CMB). The results obtained from $C_3$ are weakly sensitive to the value of $N_{\rm eff}$. The limits are the 1-$\sigma$ and the 2-$\sigma$ extremal values.}
\label{tab:bfv0}
\end{table}

\begin{table}[t]
\begin{footnotesize}\begin{center}
\renewcommand
\arraystretch{1.5}
\begin{tabular}{c|cccr|rr}
\toprule
& $\Omega_{\Lambda_0}$ & $\Omega_{\Lambda_{\infty}}$ & $H_0$ ($\text{km}\ \text{s}^{-1} \text{Mpc}^{-1}$) & \multicolumn{1}{c}{$\chi^2_{\rm min}$} & \multicolumn{1}{|c}{$\chi^2_{3}$} & \multicolumn{1}{c}{$\chi^2_{4}$} \\
\midrule
$\mu$          & 0.746 $ ^{+0.178}_{-0.122}$ $ ^{+ 0.254}_{- 0.172}$  & 
                 1.061 $ ^{+0.770}_{-0.355}$ $ ^{+01.383}_{-00.461}$  & 
                 69.94 $ ^{+00.97}_{-00.87}$ $ ^{+ 01.49}_{- 01.30}$  & 
                 565.834 & 565.878 & 566.182  \\
OHD            & 0.804 $ ^{+0.196}_{-0.570}$ $ ^{+ 0.196}_{- 0.690}$  &  
                 1.174 $ ^{+0.970}_{-0.970}$ $ ^{+01.660}_{-01.090}$  &  
                 72.68 $ ^{+07.99}_{-11.67}$ $ ^{+ 10.50}_{- 16.06}$  & 
                 7.970 & 8.824 & 8.790  \\
BAO            & 0.812 $ ^{+0.188}_{-0.775}$ $ ^{+ 0.188}_{- 0.812}$  & 
                 0.681 $ ^{+2.743}_{-0.675}$ $ ^{+40.695}_{-00.680}$  & 
                 \multicolumn{1}{c}{-}                                &  
                 0.000 & 1.185 &  1.240 \\
CMB            & 0.710 $ ^{+0.115}_{-0.189}$ $ ^{+ 0.161}_{- 0.315}$  & 
                 1.011 $ ^{+0.254}_{-0.038}$ $ ^{+01.116}_{-00.046}$  & 
                 68.89 $ ^{+14.08}_{-11.42}$ $ ^{+ 23.80}_{- 16.11}$  &
                 0.015 & \multicolumn{1}{c}{-} & 0.054\\
$C_3$          & 0.753 $ ^{+0.142}_{-0.106}$ $ ^{+ 0.229}_{- 0.153}$  &
                 1.099 $ ^{+0.549}_{-0.301}$ $ ^{+00.975}_{-00.408}$  &
                 69.97 $ ^{+00.91}_{-00.84}$ $ ^{+ 01.38}_{- 01.25}$  &
                 575.888 & \multicolumn{1}{c}{-} & \multicolumn{1}{c}{-}\\
$C_4$          & 0.720 $ ^{+0.013}_{-0.013}$ $ ^{+ 0.020}_{- 0.021}$  & 
                 1.006 $ ^{+0.020}_{-0.019}$ $ ^{+00.030}_{-00.029}$  & 
                 69.77 $ ^{+00.52}_{-00.52}$ $ ^{+ 00.79}_{- 00.79}$  &
                 576.267 & \multicolumn{1}{c}{-} & \multicolumn{1}{c}{-}\\
\bottomrule
\end{tabular}
\end{center}\end{footnotesize}
\caption{Best fit values for the interacting model parameters and the corresponding $\chi_{\rm min}^{2}$ associated with each data set and two combinations of them: $C_3$ ($\mu$+OHD+BAO) and $C_4$ ($\mu$+OHD+BAO+CMB).  The two last columns,  $\chi_{3}^{2}$ and $\chi_{4}^{2}$, represent the partial contribution of each data set to the value of $C_3$ and $C_4$, respectively. For the sake of comparison, the values obtained for the $\Lambda$CDM model using the same constraints are also shown. The limits are the 1-$\sigma$ and the 2-$\sigma$ extremal values excluding the non-physical region $\Omega_{\Lambda_0}>1$.}
\label{tab:bfv}
\end{table}

The best fit values obtained for the parameters of the $\Lambda$CDM model  and the corresponding $\chi_{\rm min}^{2}$  are shown in table~\ref{tab:bfv0}  for three different values of $N_{\rm eff}$: 3.04 , 3.14 , and 4.34. Comparing these values to those obtained from the 7-year WMAP data \cite{CMB4} ($\Omega_{\Lambda_0}=0.725\pm0.016$, $H_0=70.2\pm1.4$ km $\text{s}^{-1}\text{Mpc}^{-1}$), we find a better agreement when we use $N_{\rm eff}$=3.04. Moreover, the $\chi^2_{\rm min}$ is minimized for this value. For these reasons, we will use $N_{\rm eff}=3.04$ to constrain the interacting model.

For the $\Lambda(t)$CDM model, we present the constraints obtained from  each observational data set considered separately (distance modulus $\mu$, OHD, BAO, and CMB) as well from two different combinations of them: $C_3$ ($\mu$+OHD+BAO) and $C_4$ ($\mu$+OHD+BAO+CMB). The best fit values  and the corresponding $\chi_{\rm min}^{2}$  are shown in  in table~\ref{tab:bfv} and the projections of the  1-$\sigma$ and 2-$\sigma$ confidence regions  in the planes $\Omega_{\Lambda_0}-\Omega_{\Lambda_\infty}$, $\Omega_{\Lambda_0}-H_{0}$ and $\Omega_{\Lambda_\infty}-H_{0}$, are shown in figure~\ref{fig:confreg}. 

We consider the combination of constraints $C_3$, which excludes the CMB constraints, because there is a possible circularity problem in our analysis of the CMB data. Indeed, we have used the values of $\Omega_{b_0}$, $\Omega_{\gamma_0}$ and $\Omega_{r_0}$ obtained in the context of the $\Lambda$CDM model and we have approximated the value of $z_*$ using a fitting formula developed for the same model. In the latter case, the difficulty to obtain the exact value of $z_*$ comes from the presence of the electron density $n_e$, whose value at different redshifts depends on the recombination history, in the computation of the optical depth (eq.~(\ref{eq:OpD})). Hence, in order to have a rough idea of the error induced by this approximation, we have simply removed the electron density from eq.~(\ref{eq:OpD}) to get the integral $\int^{z_*}_{0}dz/(H(z)(1+z))$ (where $z_*$ is given by  eq.~(\ref{eq:fitz})) and compared the values obtained for the interacting and the non-interacting cases.  If the difference is not too large, that means that up to $z_*$, the evolution of the Hubble term and thus of the scale factor,  are similar in both cases; this in its turn implies that the recombination history has also to be similar since for the comparison, all the initial parameters are fixed to the same value (except $Q$). In figure~\ref{fig:aSCALE}, we see that for the CMB constraints considered alone, the percentage of difference reaches up to $\sim7.5\%$ in the 2-$\sigma$ confidence region and  $\sim5\%$ in the 1-$\sigma$ confidence region. When all the data are considered, the percentage of difference in the 2-$\sigma$ and in the 1-$\sigma$ confidence region falls respectively below $2.5\%$ and $2\%$. Hence, we can expect that eq.~(\ref{eq:fitz}) provides a reasonable approximation for $z_*$, at least in the confidence regions obtained from the combination of constraints $C_4$.

We can notice that the confidence regions obtained from BAO (1-$\sigma$) , from OHD (1-$\sigma$)  and from the distance modulus $\mu$ (2-$\sigma$) extend partially beyond $\Omega_{\Lambda_0}=1$. However, a value of $\Omega_{\Lambda_0}$ greater than 1 would imply a negative value for the current energy density of matter ($\rho_{m_0}<0$) and therefore must be considered as non-physical. 
Similarly, we can also notice that the 1-$\sigma$ confidence regions obtained from BAO and OHD extend (marginally in the case of OHD) into region of the parameter space where $k>0$ (below the dot-dashed line on figure \ref{fig:confreg}). We have shown in section~ \ref{sec:MDL} that the solutions with $k>0$  necessarily reach a point where $\rho_{m}$ becomes negative if we are looking sufficiently far in the past. However, for the redshifts involved in the computation of the BAO and the OHD constraints, this point is never reached and $\rho_{m}$ remains always positive. Hence, in this case, there is no reason to consider these constraints as non-physical. In any case, these two regions  are excluded by at least 2-$\sigma$ from the combinations of constraints $C_3$ and $C_4$.

Regarding the best fit values obtained, the processes of decay of dark matter into dark energy ($\Omega_{\Lambda_{\infty}}>1$) is favoured by only one constraint (BAO), while the decay of dark energy into dark matter ($\Omega_{\Lambda_{\infty}}<1$) is favoured by the three other, as well as by the combinations $C_3$ and $C_4$. However, all the 1-$\sigma$ confidence regions extend below and above the the plane $\Omega_{\Lambda_{\infty}}=1$ (which corresponds to the $\Lambda$CDM model) and therefore both processes still are statistically allowed.

As shown in section~\ref{sec:MDL}, we need to have $\Omega_{\Lambda_{\infty}}\approx\Omega_{\Lambda_{0}}$ in order to relate and explain the cosmological constant problem and the  coincidence problem. However, as we can see on figure~\ref{fig:confreg}, these values are excluded by at least 2-$\sigma$ due to constraints $C_4$. Actually, the closest point to the line $\Omega_{\Lambda_{\infty}}=\Omega_{\Lambda_{0}}$  situated on the 2-$\sigma$ contour is $\Omega_{\Lambda_0}=0.718$,  $\Omega_{\Lambda_{\infty}}=0.983$. For these parameters,  the value of $\rho_{\Lambda}$ remains mostly constant ($\rho_{\Lambda_i}/\rho_{\Lambda_0}\approx1.04$), and a coincidence problem is still  present since the current ratio of matter and dark energy density ($\rho_{m_{0}}/\rho_{\Lambda_0}=0.393$) lies in the transition zone between the point where  dark energy starts to dominate and the late-time phase where the ratio $\rho_{m}/\rho_{\Lambda}$ becomes approximately constant ($\rho_{m_{\infty}}/\rho_{\Lambda_{\infty}}=0.017$).

The best fit values obtained for the $\Lambda$CDM model from the constraints $C_4$ (table~\ref{tab:bfv0}) are included in the 1-$\sigma$ confidence region of the interacting model. Moreover, the $\chi^2_{\rm min}$ is only slightly smaller for interacting model (576.267 vs. 576.638). Hence, we may wonder if the introduction of an extra-parameter ($\Omega_{\Lambda_{\infty}}$) is really justified. The improvement of the $\chi^2_{\rm min}$ value may be assessed by the mean of the Bayesian information criterion \cite{BIC1}, defined as
\begin{equation}
   \text{BIC}=-2\ln\mathcal{L}_{\rm max} + K\ln N=(\chi^{2}_{\rm min} + C) + K\ln N,
   \label{eq:BIC}
\end{equation}
where $\mathcal{L}_{\rm max}$ is the maximum likelihood, $K$ is the number of parameters for the model (2, for the $\Lambda$CDM model, 3 for the interacting model), $N$ the number of data points used in the fit ($N=635$) and $C$ a constant independent of the model used. Following ref.~\cite{BIC2}, we will regard a difference of 2 for the BIC as a non significant, and of 6 or more as very non-significant improvement of the $\chi^2_{\rm min}$ value. Since we get $\Delta_{\text{BIC}}=6.08$, we can conclude that the addition of an extra parameter is not warranted by the marginal decrease in the value of $\chi^2_{\rm min}$. 

If we consider now the combination of observational constraints $C_3$ (which excludes the CMB data), the improvement of the $\chi^2_{\rm min}$ value remains insignificant  ($\Delta_{\text{BIC}}$=6.15) and the closest point to the line $\Omega_{\Lambda_{\infty}}=\Omega_{\Lambda_{0}}$ on the 2-$\sigma$ contour is now $\Omega_{\Lambda_0}=0.615$ , $\Omega_{\Lambda_{\infty}}=0.693$, which leads to the ratio $\rho_{\Lambda_i}/\rho_{\Lambda_0}\approx3.37$,  still far to being able to explain the discrepancy with the theoretical value.  The current value of the ratio $\rho_{m}/\rho_{\Lambda}$  is now closer to the late-time value (0.443 vs. 0.626), but not sufficiently to explain the coincidence problem.

\subsection{General case}

\begin{figure}[ht]
  \centering
  \includegraphics[scale=0.75]{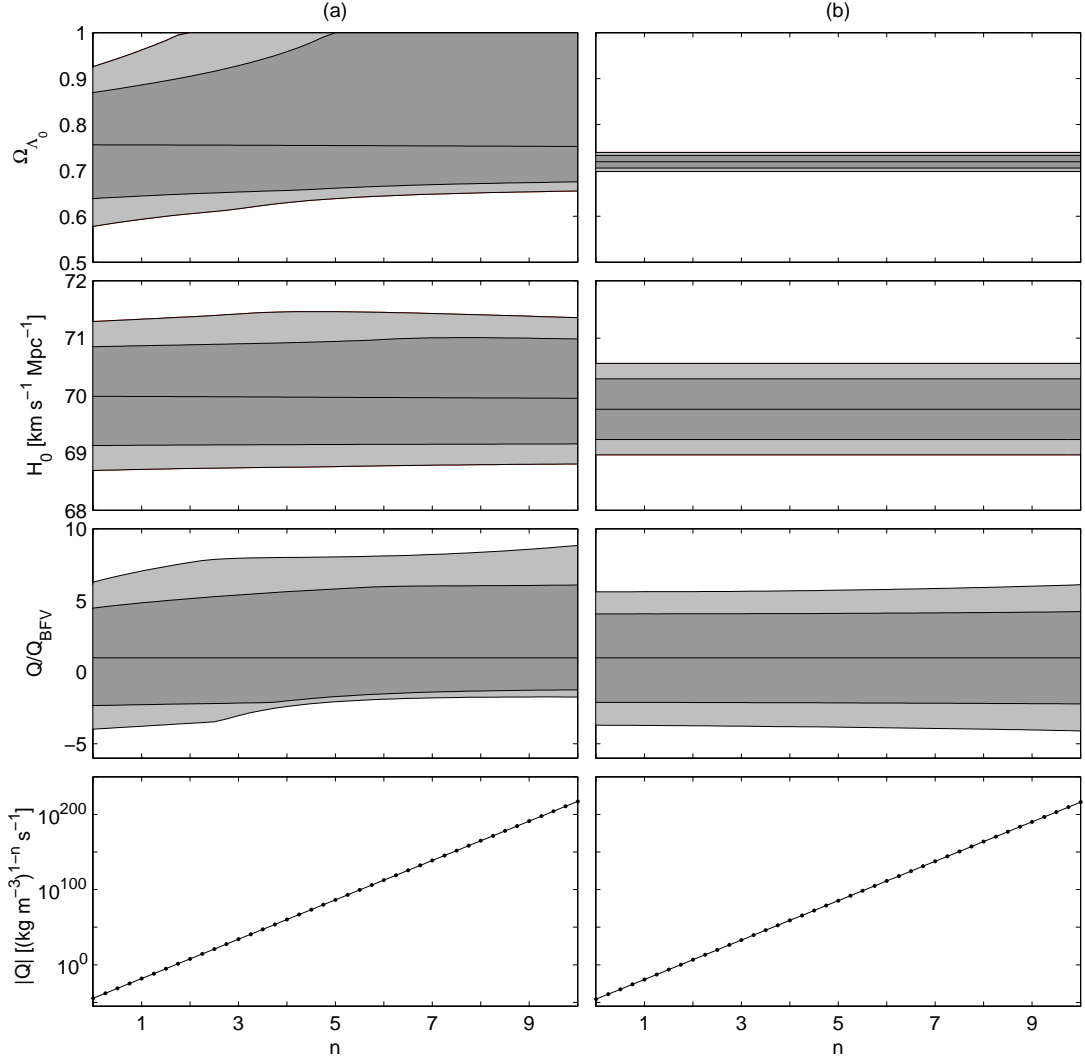}
  \caption{The panels in the first three rows show the best fit values (solid line), the 1-$\sigma$ (dark shade) and the 2-$\sigma$ (light shade) confidence intervals for $\Omega_{\Lambda_0}$, $H_0$ and $Q/Q_{BFV}$ at different value of $n$. In the panels of the last row, the dots represents the best fit values $Q_{BFV}$ and the solid line is given by eq.~(\ref{eq:Qn}) where we have used the best fit value found in section~\ref{sect:n32} ($\tilde{n}=3/2)$ for $\tilde{Q}$ and $\rho_{\Lambda_0}$. For $Q/Q_{BFV}$, since $Q_{BFV}$ turns out to be always negative, $Q/Q_{BFV}>0$ corresponds to  dark matter decay and  $Q/Q_{BFV}<0$ to dark energy decay. These results were obtained using the combination of constraints $C_3$ ($\mu$+OHD+BAO)  in (a), and the combination of constraints $C_4$ ($\mu$+OHD+BAO+CMB) in (b). }
  \label{fig:genn}
\end{figure}

The motivation to study the model with $n=3/2$ was purely phenomenological and  did not rely upon any deep principles. Then, it will be interesting to extend our analysis to general values of $n$. However, we do not perform an analysis as detailed as for $n=3/2$ and restrict it to constrain the model parameters in the range $n=0$ to $n=10$. The results are presented in figure~\ref{fig:genn} for the combination of observational constraints $C_3$ and $C_4$.

We can see on figure~\ref{fig:genn} that the best fit values for $\Omega_{\Lambda_0}$ and $H_0$ are nearly constant for both $C_3$ and $C_4$, and for the latter, this is even the case for the confidence intervals.  These results seem a priori very surprising, but could be understand if the constant $Q$ is sufficiently small (though still non-zero). Indeed, in the limit of a small $Q$, it is possible to reproduce exactly the same cosmological evolution, for $t<t_0$, using two different values of $n$. 

If we suppose that the term $Q\rho_{\Lambda}^{n}$ can be neglected in the matter energy conservation equation (eq.~(\ref{eq:conmat})),  we find a dependence on $n$ only in the energy density $\rho_{\Lambda}$ and in the quantities derived from it. We are then interested to determined the conditions needed to have $\rho_{\Lambda_{n}}(t)=\rho_{\Lambda_{\tilde{n}}}(t)$, with $n\neq\tilde{n}$. In the limit where $Q$ is small, the solutions found for $\rho_{\Lambda}$ (eq.~(\ref{eq:convac})) become
\begin{equation}
\rho_{\Lambda_{n}}(t)=\rho_{\Lambda_0}-Q\rho_{\Lambda_0}^{n}t.
  \label{eq:denvacsq}
\end{equation}
Hence the condition to have $\rho_{\Lambda_{n}}(t)=\rho_{\Lambda_{\tilde{n}}}(t)$ is given by 
\begin{equation}
  Q=\tilde{Q}\rho_{\Lambda_0}^{\tilde{n}-n},
  \label{eq:Qn}
\end{equation}
which provides an explanation to the exponential relationship observed on the lower panels of figure~\ref{fig:genn} between the best fit value of $Q$ and $n$.

For the best-fit values, the same cosmological evolution is closely reproduced independently of the value of $n$, hence we find the same value for $\chi^2_{\rm min}$. Concerning the 1-$\sigma$ and 2-$\sigma$ confidence regions, the results are weakly sensitive to $n$ for the constraints $C_4$, (for the higher values of $n$, $Q$ becomes sufficiently large to start to observe a weak deviation from the exponential relationship) but are more much important for the constraints $C_3$. However, even in this case, the constraints obtained do not allow ${\rho_{\Lambda}}$ to decay from an initially large value to that observed today.

  \section{Conclusion}

In this paper, motivated by the hope to shed some light on the coincidence and the cosmological constant problems, we have applied the phenomenological decay law $\dot{\rho}_{\Lambda}=-Q\rho_{\Lambda}^{n}$, originally proposed in ref.~\cite{Hiscock} to described the decay of dark energy into radiation, to study the interaction between the dark sectors. In order to ameliorate these two problems, we were primarily interested in the decay of dark energy into dark matter ($Q>0$), but we have also considered the decay of dark matter into dark energy ($Q<0$).

From  dimensional analysis, we have shown that the model with $n=3/2$ (and only this model) leads to solution where the ratio of energy densities  $\rho_{m}/\rho_{\Lambda}$ is constant at late time.  An important feature of this model is the possibility to have, in addition to this late phase (which could explain the coincidence problem), an early phase  qualitatively similar to what is obtained in the $\Lambda$CDM model (which already provides a good fit to the observational constraints).

To constrain the three free parameters of the model with $n=3/2$, i.e. the current density parameter of dark energy ($\Omega_{\Lambda_0}$), the late-time density parameter of dark energy ($\Omega_{\Lambda_\infty}$ $-$ related to $Q$) and the current Hubble parameter ($H_0$), we have mainly followed the procedure of  refs.~\cite{DPZ,XuLu,TongNoh,LuXuWu} and considered the observational constraints involving the distance modulus $\mu$ of type Ia supernova (SNeIa) and  gamma-ray bursts (GRB), the baryon acoustic oscillation (BAO), the cosmic microwave background (CMB) and the observational Hubble rate (OHD). The constraints obtained from the CMB data could be affected by a circularity problem. To  constrain the $\Lambda(t)$CDM model with the CMB data,  we have used some parameters obtained from the usual $\Lambda$CDM model  ($\Omega_{b_0}$, $\Omega_{\gamma_0}$,  $\Omega_{r_0}$,  $z_*$). Although we have argued that it was legitimate to do so, that could possibly lead to biased results. For this reason, we have presented our result considering two different combinations of observational constraints, $C_3$ ($\mu$+OHD+BAO) and $C_4$ ($\mu$+OHD+BAO+CMB).

For the constraints $C_4$, the best fit parameters  are given by $\Omega_{\Lambda_0}=0.720$, $\Omega_{\Lambda_\infty}=1.006$ and $H_0=69.77 \text{ km}\ \text{s}^{-1} \text{Mpc}^{-1}$. Since $\Omega_{\Lambda_\infty}>1$ corresponds to $Q<0$ and $\Omega_{\Lambda_\infty}<1$ to $Q>0$, the process involved at the best fit point is the decay of dark matter into dark energy and in fact exacerbates the coincidence and the cosmological constant problems. We have shown that in order to explain both the coincidence and the cosmological constant problems, we need to have $\Omega_{\Lambda_0}\approx\Omega_{\Lambda_{\infty}}$. The point lying in the 2-$\sigma$ confidence region where these parameters are closest is $\Omega_{\Lambda_0}=0.718$,  $\Omega_{\Lambda_{\infty}}=0.983$. For these values, $\rho_{\Lambda}$ remains nearly constant ($\rho_{\Lambda_i}/\rho_{\Lambda_0}\approx1.04$) and the current value of the ratio $\rho_{m}/\rho_{\Lambda}$ is far from the late-time value (0.393 vs. 0.017).

For the constraints $C_3$, the best fit parameters ($\Omega_{\Lambda_0}=0.753$, $\Omega_{\Lambda_\infty}=1.099$ and $H_0=69.97 \text{ km}\ \text{s}^{-1} \text{Mpc}^{-1}$) also  correspond to the decay of dark matter into dark energy. The closest point (in the 2-$\sigma$ confidence region) to the line $\Omega_{\Lambda_0}=\Omega_{\Lambda_{\infty}}$ is now $\Omega_{\Lambda_0}=0.615$, $\Omega_{\Lambda_{\infty}}=0.693$. At this point, the value of  $\rho_{\Lambda_i}/\rho_{\Lambda_0}$ remains far away from the desired value ($\approx3.37$), but the current and late-time value of the ratio  $\rho_{m}/\rho_{\Lambda}$  are now of the same order of magnitude (0.626 and 0.443, respectively). However, to provide a convincing explanation to the coincidence problem, we should have $\rho_{m_{0}}/\rho_{\Lambda_{0}}\approx\rho_{m_{\infty}}/\rho_{\Lambda_{\infty}}$, which is not the case here.

From a statistical point of view, we do not find that the addition of an extra parameter was justified. The $\chi_{\rm min}^{2}$ values obtained for the $\Lambda(t)$CDM model were only slightly smaller than the value obtained for the $\Lambda$CDM, 575.888 vs. 576.178 for $C_3$ and 576.267 vs. 576.638 for $C_4$. In both cases, according to the Bayesian information criterion (BIC), the improvement of the $\chi_{\rm min}^{2}$ is considered to be insignificant ($\Delta_{BIC}=6.15$ for 
$C_3$ and $\Delta_{BIC}=6.08$  for $C_4$).

Finally, we have extended our results to generic values of $n$ ranging form 0 to 10. The values of $\chi^2_{\rm min}$ and the corresponding values for $\Omega_{\Lambda_0}$ and for $H_0$ are nearly constant, whereas the value of $Q$ have a exponential dependence on $n$. We have shown that these results may be understood in the limit of a small $Q$. Since the same values of $\chi^2_{\rm min}$ are found, the conclusion that we drew for the BIC holds for any value of $n$, i.e. the improvement of the $\chi_{\rm min}^{2}$ is insignificant. Moreover, even if we consider the 2$\sigma$-confidence region, the constraints obtained do not offer any explanation for the cosmological constant problem.

Since when we consider the exclusion limits obtained from observational data, the model studied in this article fails to provide an explanation to the coincidence and the cosmological constant problems, and since this model is also disfavoured by the BIC, we can conclude that the usual $\Lambda$CDM model remains the most reasonable description of the Universe.

\clearpage
\noindent{\bf Acknowledgement}\\

\vspace{0mm}

We would like to thank James Cline for helpful comments on the preliminary version of this article. This work has been supported by the Fonds de recherche du Qu\'ebec - Nature et technologies (FQRNT) through its doctoral research scholarships programme.

\end{document}